\documentclass[prb,twocolumn,letterpaper,showpacs,preprintnumbers,superscriptaddress,amsmath,amssymb]{revtex4}

\usepackage{graphicx}% Include figure files
\usepackage{dcolumn}% Align table columns on decimal point
\usepackage{bm}% bold math

\begin{document}

\title{Acoustoelectric current transport through single-walled carbon nanotubes}

\author{J. Ebbecke}
 \email{jens.ebbecke@physik.uni-augsburg.de}
\affiliation{Institut f{\"u}r Physik der Universit{\"a}t Augsburg, Experimentalphysik I,  Universit{\"a}tsstr. 1, 86135 Augsburg, Germany}
\author{C. J. Strobl}
\affiliation{Sektion Physik der Ludwig-Maximilians-Universit{\"a}t and Center for NanoScience (CeNS), Geschwister-Scholl-Platz 1, 80539 Munich, Germany}
\author{A. Wixforth}
\affiliation{Institut f{\"u}r Physik der Universit{\"a}t Augsburg, Experimentalphysik I,  Universit{\"a}tsstr. 1, 86135 Augsburg, Germany}

\date{\today}% It is always \today, today, but any date may be explicitly specified

\begin{abstract}
We have contacted single-walled carbon nanotubes after aligning the tubes by the use of surface acoustic waves. The acoustoelectric current has been measured at 4.2 K and a probing of the low-dimensional electronic states by the surface acoustic wave has been detected. By decreasing the acoustic wavelength resulting in an adjustment to the length of the defined carbon nanotube constriction a quantization of the acoustoelectric current has been observed. 
\end{abstract}


\maketitle

Due to their remarkable electrical and mechanical properties carbon nanotubes (CNT) have become an important field of research since their discovery in 1991 \cite{Iijima91}. They can be synthesized as single-walled carbon nanotubes (SWNT) as well as multi-walled carbon nanotubes (MWNT). Because of their small diameter of around 1 nm and a length of up to several micrometers SWNTs are by definition an ideal one-dimensional electronic system. Efforts have been made to investigate fundamental one-dimensional effects like Luttinger liquid behaviour \cite{Egger99}. Also by contacting the CNTs with metal leads zero-dimensional electronic systems can be designed \cite{Bockrath97,Tans97,Tans98,Yao99,Ida00,Suzuki01,Park02}. These quantum dots (QD) with a length between micrometers and several tens of nanometers have shown an increased zero-dimensional confinement in comparision with their semiconductor counterparts of the same length.

In this manuscript, we report on the observation of an acoustoelectric current transport through carbon nanotube constrictions (CNTC). For large wavelengths of the surface acoustic wave (SAW) the current oscillates by increasing the SAW amplitude that exhibits a probing of the low-dimensional electronic states. By adjusting the acoustic wavelength to the length of the CNTC the acoustoelectric current becomes quantized for certain SAW amplitudes. At first sight, these results resemble the proposed quantized adiabatic charge transport in CNTs of Ref.\onlinecite{Talyanskii01}. There, the potential of the SAW is assumed to induce a miniband spectrum in the CNTs and an adiabatic quantized transport as described by Thouless\cite{Thouless83} is suggested. But in contrast to this mechanism here a turnstile like modulation of the tunneling barriers by the piezoelectric potential of the SAW leads to the acoustoelectric current quantization. Therefore this work is related to the experiments of Ref.\onlinecite{Ebbecke04} where a single electron transport by surface acoustic waves through a QD induced in a AlGaAs/GaAs heterostructure has been presented. Also beside the fact that Coulomb repulsion is likely to be important for the current quantisation the results presented in this paper are in close relation to the proposed carbon nanotube electron pump\cite{Wei01}.

\begin{figure}
  \begin{center}
	 \includegraphics[width=3in]{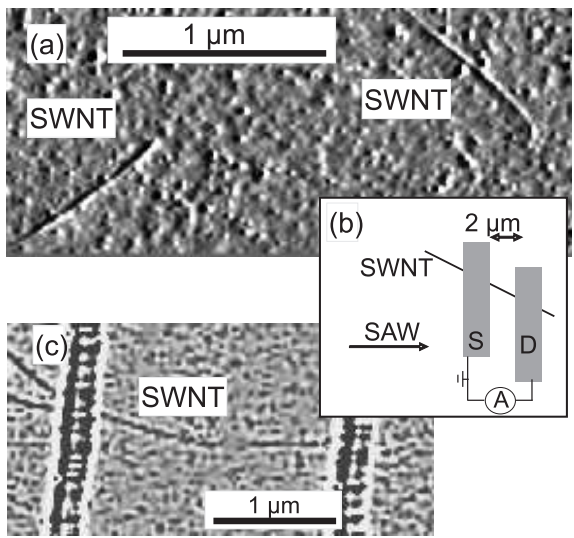}
	\end{center} 
	\caption{(a) AFM picture of two short SWNTs aligned by the SAW at $\pm$45$^\circ$. (b) schematic sample layout (c) an AFM picture of a contacted SWNT of sample A}
	\label{fig:fig1}
\end{figure}

Two sets of samples have been processed almost solely by optical lithography and results will be presented exemplarily for one sample of each set (in the following named sample A and sample B). The only difference is the periodicity of the interdigital transducers (IDT). On sample A first two IDTs with a periodicity of 35~$\mu$m have been processed on LiNbO$_3$ substrate (rotation 128$^\circ$ Y-cut X-propagation) by optical lithography. At a sound velocity of the SAW of $v$ = 3850 m/s (at $T$ = 4.2 K) the resonance frequency of the transducers is given by $f_1$ = $v/p$ = 110 MHz. On sample B two IDTs have been defined by ebeam lithography with a periodicity of 8.1~$\mu$m resulting in a primary frequency of $f_1$ = 475 MHz. SWNTs grown by an arc discharge method have been suspended in water with 1 wt$\%$ sodium-dodecylsulfate (SDS). The suspension was subjected to ultrasonic agitation for 10 minutes and then centrifuged at 10,000 g for 10 minutes to remove larger particles. Small drops of the liquid have been dispensed between the transducers and glass plates have been placed above. A SAW launched with one of the IDTs has two effects. First the parallel component of its piezoelectric field aligns the NTs in parallel to the wave vector of the SAW. Secondly the transversal component of the crystal particle movement induces strong fluidic processes in the liquid. The superposition of both effects is an alignment of the NTs with an angle between 25$^\circ$ and 45$^\circ$ with respect to the propagation direction of the SAW. This process of aligning NTs has been described in detail in Ref.\onlinecite{Strobl04} where MWNTs have been aligned. In Fig.~\ref{fig:fig1} (a) an atomic force microscope (AFM) picture of short aligned SWNTs is shown. The nanotubes are aligned with an angle of $\pm$45$^\circ$ with respect to the travelling wave. After cleaning the samples by rinsing them in deionized water and drying with nitrogen gas pairs of metal contacts have been processed where the drops of SWNT suspension had been (see schematic sample layout given in Fig.~\ref{fig:fig1}(b) ). The contacts are metal fingers of 30 nm thick titanium with a length of 600~$\mu$m and a width of 2~$\mu$m. The gap between the contact pair is also 2~$\mu$m so that the contacted SWNTs have a length between 2.2~$\mu$m and 2.8~$\mu$m depending on the alignment angle between 25$^\circ$ and 45$^\circ$. An AFM picture of a contacted SWNT of sample A is shown in Fig.~\ref{fig:fig1} (c). All measurements have been carried out in a variable temperature cryostate with base temperature of $T$ = 4.2 K.

\begin{figure}
  \begin{center}
	 \includegraphics[width=3in]{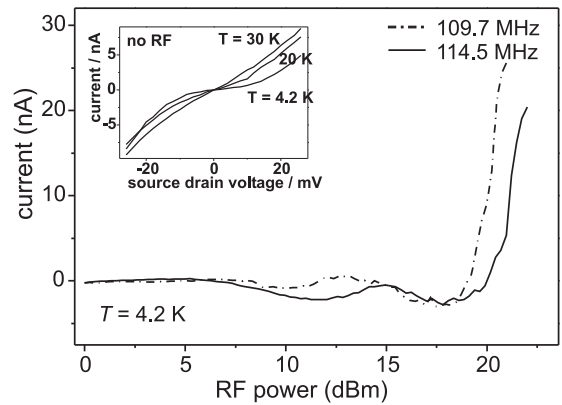}
	\end{center} 
	\caption{Sample A: acoustoelectric current oscillations by raising the applied RF power for two frequencies inside the passband of the IDT. Source-drain measurements for different temperatures are shown as an inset.}
	\label{fig:fig2}
\end{figure}

First, investigations made with sample A will be presented. The two-terminal resistance of the device at room temperature was 500 k$\Omega$. By using the sample holder as a back gate no change in conductance has been detected for voltages of $V_{BG}$=$\pm$20 V at room temperature (also for sample B). Consequently we assume the contacted CNTs to be metallic. The current transport through the SWNT has been detected using a Keithley 2400 SourceMeter as a function of applied source drain voltage for temperatures between $T$ = 4.2 K and $T$ = 30 K (inset of Fig.~\ref{fig:fig2}). At $T$ = 4.2 K a clear nonlinearity has been measured that has almost disappeared at $T$ = 30 K. The espected Coulomb blockade behaviour is masked by the vanishing conduction of the contacts to the CNT and the capacitance of the back gate is to small to be used in this case. In Ref.\onlinecite{Suzuki01} a value of the charging energy of a CNT quantum dot with 1.3~$\mu$m length has been measured to be $E_C$ = 6 meV. With the estimation of a linear dependence of the capacity on the length of the CNT a value of $E_C~\approx~$2 meV can be assumed for the CNTCs presented here. Due to the fact that we can not verify any Coulomb oscillations because of the lack of a gate and also that a 2~$\mu$m long constriction with a diameter of a few nanometers is venturous to term a zero-dimensional quantum dot we name our devices CNT constrictions but still in mind that the CNTC have well-defined electronic states.

Two measurements of the acoustoelectric current as a function of applied RF power are shown in Fig.~\ref{fig:fig2} for two different frequencies inside the passband of the IDTs of sample A (using a Keithley 617 electrometer and a RohdeSchwarz signal generator SMP2). The piezoelectric field of the SAW transmits a momentum to the electrons in the SWNT and an acoustoelectric current can be detected. By raising the RF power meaning an increase of the SAW amplitude the acoustoelectric current through the CNTC is oscillating for both frequencies used. The horizontal shift in RF power of the current minima are caused by the fact that the transducer is effective differently in launching a SAW at these two frequencies. After finishing the measurements and taking the figure of a contacted CNT (see Fig.~\ref{fig:fig1}(c) ) we placed sample A for 30 s into an oxigen plasma. No conductivity of sample A has been detected afterwards.

Similar oscillations of the acoustoelectric current have also been detected in AlGaAs/GaAs heterostructures\cite{Shilton96,Talyanskii98,Fletcher03}. In split-gate induced one-dimensional channels unintentional QDs can be formed accidently by the potential of impurities. Pronounced current oscillations have been measured close to conduction pinch off in these samples. In Ref.\onlinecite{Shilton96} the assumption has been made that the rather complicated charge transport could be dominated by Coulomb blockade type effects. This idea has been investigated in more detail in Ref.\onlinecite{Fletcher03} where also an unintentional QD was present in the induced one-dimensional channel of a GaAlAs/GaAs heterostructure. It has been shown that the potential of the SAW modulates the tunnel barriers of the accidently confined QD and a SAW mediated tunneling of electrons through the zero-dimensional electronic states has been proposed. If the SAW wavelength exceeds the length of the confined low-dimensional potential (like sample A with $\lambda_{SAW}$ = 35~$\mu$m) also an acousto-electric current of reversed sign has been detected\cite{Shilton96,Talyanskii98} (and in Fig.~\ref{fig:fig2}). The piezoelectric potential is supposed to lower both the potential barriers at the entrance and the exit of the constriction and dependent of the relative position of the electronic states in the constriction on the Fermi levels in the metal contacts a resulting acoustic-electric current with both signs has been detected. 

A continuation of the work of Ref.\onlinecite{Fletcher03} has been presented recently\cite{Ebbecke04}. A QD has been induced deliberately by three independent metallic split gates in a AlGaAs/GaAs heterostructure and a single electron transport by SAWs has been demonstrated. The length of the lithographically defined QD was half the wavelength of the SAW and the transport mechanism is a turnstile like tunneling of electrons through the QD.

\begin{figure}
  \begin{center}
	 \includegraphics[width=3in]{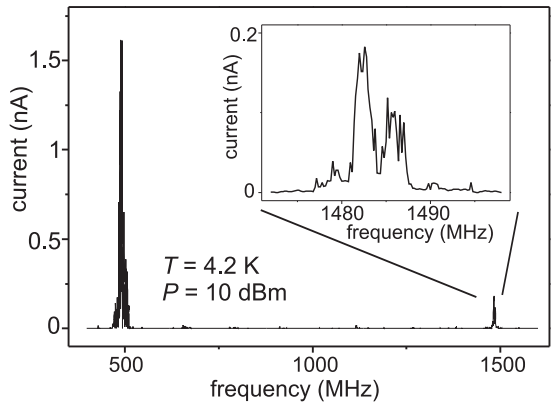}
	\end{center} 
	\caption{Sample B: acoustoelectric current as a function of applied RF frequency. The inset shows an enlargement of the current at the frequency of the third harmonic.}
	\label{fig:fig3}
\end{figure}

In the following, experiments made with sample B will be presented. IDTs have been processed with a periodicity of 8.1~$\mu$m meaning an adjustment of the acoustic wavelength to the length of the CNTC.
The acoustoelectric current transport through a SWNT as a function of applied RF frequency is shown in Fig.~\ref{fig:fig3}. Around the resonance frequency $f_1$ = 475 MHz an acoustoelectric current has been measured. But in addition to this signal at $f_1$ another peak of acoustoelectric current exists at a frequency of approximately $f_3$ = 1480 MHz that is slightly higher than the theoretical value of the third harmonic $f_{3theo}$ = 1425 MHz. This SAW launched by the same IDT has a wavelength of approximately 2.6~$\mu$m. The two-terminal resistance of sample B at room temperature was 6.5 M$\Omega$ and therefore significantly higher than the resistance of sample A. This increase in resistance is caused by less effective electrical contacts to the SWNT resulting in a suppression of current transport at $T$ = 4.2 K in the interval of $\pm$2 V source drain bias. Due to the lack of any appropriate gate (like in sample A) the charging energy of the CNTC could not be determined.

\begin{figure}
  \begin{center}
	 \includegraphics[width=3in]{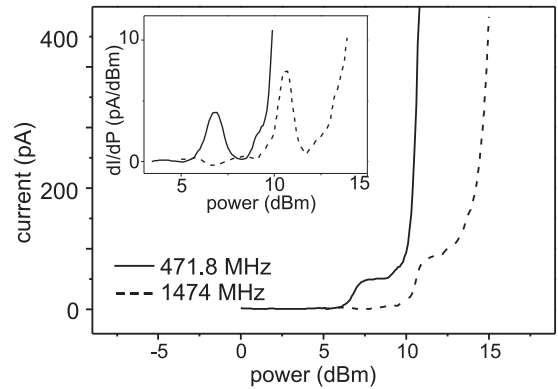}
	\end{center} 
	\caption{Quantized acoustoelectric current as a function of applied RF power for two different SAW frequencies (first and third harmonic of the IDT) at 4.2 K. The upper graph shows the derivative of the current.}
	\label{fig:fig4}
\end{figure}

In Fig.~\ref{fig:fig4} two measurements of the acoustoelectric current as a function of applied RF frequency are shown. The solid graph has been measured at a frequency inside the passband of the IDT's first harmonic and the other graph with a frequency of the IDT's third harmonic. For both frequencies the acoustoelectric current is zero for low RF power levels. By raising the RF power resulting in a launched SAW with larger amplitude the current increases until a plateau is reached. As a verification of the occurrence of plateaus the numerical derivative of the current is shown in the inset of Fig.~\ref{fig:fig4} where the pronounced minima indicate the plateaus. At these RF powers levels the acoustoelectric current is quantized. The RF levels at which plateaus occur are different for the two frequencies because the IDT is effective differently in launching the first and third harmonic (see Fig.~\ref{fig:fig3}). By further increasing the RF power the current also increase without showing any further well quantized levels. A feature of this current quantization remaining to be explained is the value of the current plateaus. For the higher frequency the plateau value is $I$ = 91 pA in contrast to the expected value of $I_3=ef_3$ = 236 pA and for the lower frequency the quantized value is $I$ = 49 pA where the theoretical current should be $I_1=ef_1$ = 75 pA. Although the current value on the shoulder visible next to the plateau for $f$ = 471.8 MHz (also visible as a shoulder in the derivative in the inset of Fig.~\ref{fig:fig4}) is $I$ = 76 pA $\approx$ $ef_1$ no pronounced current quantization at $I=ef$ has been detected.

\begin{figure}
  \begin{center}
	 \includegraphics[width=3in]{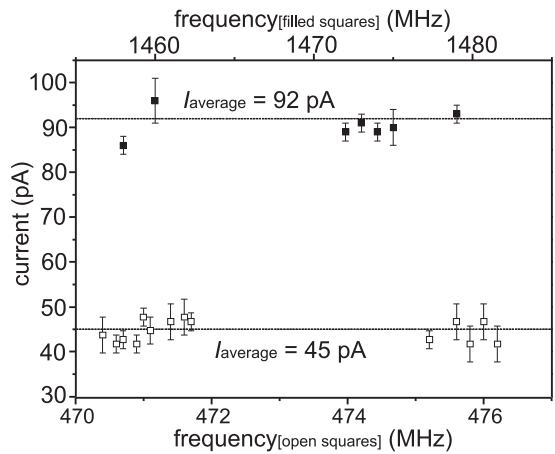}
	\end{center} 
	\caption{Statistical overview of the height of the first current plateaus as a function of frequency.}
	\label{fig:fig5}
\end{figure}

In order to emphasize that these quantized acousto-electric current values are not randomly a statistic of all measured first plateau current values are given in Fig.~\ref{fig:fig5}. The values of both the first and the third harmonic are shown in the same figure. The open squares are belonging to the lower frequency $f_1$ and their average value is $I_{average}$ = 45 pA that is 60 $\%$ of $I=ef_1$. The filled squares are associated with the third harmonic $f_3$ with an average current of $I_{average}$ = 91 pA that is 40 $\%$ of $I=ef_3$. A similar "fractional" plateau has been presented in Ref.\onlinecite{Fletcher03} for quantized acoustoelectric current transport measurements through an unintentional quantum dot defined in a AlGaAs/GaAs heterostructure. We assume that some of the electrons are excited during the transport mechanism and the raised tunneling probability of electrons from the excited states of the CNTC back to the source contact could be the reason for the lowered plateau height. These "`fractional"' cannot be explained by a parallel transport through two or more CNTs because is has been shown that a parallel transport results in an increased plateau height\cite{Ebbecke00}.

In Fig.~\ref{fig:fig5} it can also be seen that there are frequencies inside the passband of the IDTs where no current quantisation has been detected (e.g. between $f$ = 472 MHz and $f$ = 475 MHz). This is caused by reflected SAWs from the second unused IDT and has been investigated in detail in Ref.\onlinecite{Cunningham99}. There in agreement with Fig.~\ref{fig:fig3} also no current quantisation as a function of rf frequency has been detected.

\begin{figure}
  \begin{center}
	 \includegraphics[width=3in]{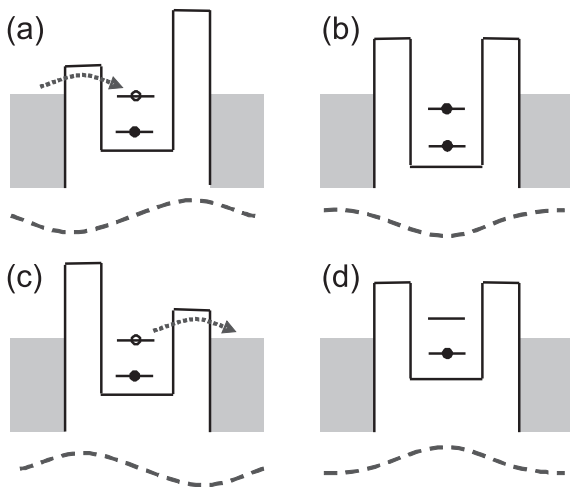}
	\end{center} 
	\caption{Schematic model of the transport mechanism. One by one electron is cycled through the CNTC by the piezoelectric SAW potential (sketched as dashed lines).}
	\label{fig:fig6}
\end{figure}

A schematic model of the assumed transport mechanism in conformance with the results presented in Ref.\onlinecite{Ebbecke04} is sketched in Fig.~\ref{fig:fig6}. First the piezoelectric potential of the SAW lowers the tunneling barrier of the metal/CNTC contact and an additional electron is tunneling onto the CNTC. Half a period later (see Fig.~\ref{fig:fig6}(c) ) the exit potential barrier is lowered and the electron is leaving the CNTC towards the drain contacts. This turnstile transport mechanism leads to a quantized current. The presented results of a SAW mediated SWNT electron pump is in general similar to the theoretical work of a CNT-based parametric electron pump\cite{Wei01} except the fact that Coulomb repulsion is essential in the turnstile model.

In summary we have realised an acoustoelectric current transport through SWNTs. For larger wavelength the acoustoelectric current oscillates as a function of applied RF power showing a probing of the low-dimensional electronic states by the potential of the SAW. By decreasing the wavelength of the SAW a quantization of the acoustoelectric current has been detected. 

\begin{acknowledgments}
The authors like to thank C. Dupraz and U. Beierlein for their support with the SWNTs. The work on the NT alignment was funded in part by the Bayrische Forschungsstiftung under the program ForNano.
\end{acknowledgments}


\begin{thebibliography}{04}

\bibitem{Iijima91} S. Iijima, Nature 354, 56 (1991)
\bibitem{Egger99} R. Egger, Phys. Rev. Lett. 83, 5547 (1999)
\bibitem{Bockrath97} M. Bockrath, D. H. Cobden, P. L. McEuen, N. G. Chopra, A. Zettl, A. Thess, and R. E. Smalley, Science 275, 1922 (1997)
\bibitem{Tans97} S. J. Tans, M. H. Devoret, H. Dai, A. Thess, R. E. Smalley, L. J. Geerligs, and C. Dekker, Nature 386, 474 (1997)
\bibitem{Tans98} S. J. Tans, M. H. Devoret, R. J. A. Groeneveld, and C. Dekker, Nature 394, 761 (1998)
\bibitem{Yao99} Z. Yao, H. W. Ch. Postma, L. Balents, and C. Dekker, Nature 402, 273 (1999)
\bibitem{Ida00} T. Ida, K. Ishibashi, K. Tsukagoshi, Y. Aoyagi, and B. W. Alphenaar, Superlattices and Microstructures 27, 551 (2000)
\bibitem{Suzuki01} M. Suzuki, K. Ishibashi, T. Ida, D. Tsuya, K. Toratani, and Y. Aoyagi, J. Vac. Sci. Technol. B 19, 2770 (2001)
\bibitem{Park02} J. W. Park, J. B. Choi, and K.-H. Yoo, Appl. Phys. Lett. 81, 2644 (2002)
\bibitem{Talyanskii01} V. I. Talyanskii, D. S. Novikov, B. D. Simons, and L. S. Levitov, Phys. Rev. Lett. 87, 276802 (2001)
\bibitem{Ebbecke04} J. Ebbecke, N. E. Fletcher, T. J. B. M. Janssen, F. J. Ahlers, M. Pepper, H. E. Beere, and D. A. Ritchie, Appl. Phys. Lett. 84, 4319 (2004)
\bibitem{Strobl04} C. J. Strobl, C. Sch{\"a}flein, U. Beierlein, J. Ebbecke, and A. Wixforth, Appl. Phys. Lett. 85, (2004)
\bibitem{Shilton96} J. M. Shilton, D. R. Mace, V. I. Talyanskii, Y. Galperin, M. Y. Simmons, M. Pepper, and D. A. Ritchie, Journal of Physics - Condensed Matter 8, L337 (1996)
\bibitem{Fletcher03} N. E. Fletcher, J. Ebbecke, T. J. B. M. Janssen, F. J. Ahlers, M. Pepper, H. E. Beere, and D. A. Ritchie, Phys. Rev. B 68, 245310 (2003)
\bibitem{Talyanskii98} V. I. Talyanskii, J. M. Shilton, J. Cunningham, M. Pepper, C. J. B. Ford, C. G. Smith, E. H. Linfield, D. A. Ritchie, and G. A. C. Jones, Physica B 249-251, 140 (1998)
\bibitem{Ebbecke00} J. Ebbecke, G. Bastian, M. Bl{\"o}cker, K. Pierz, and F. J. Ahlers, Appl. Phys. Lett 77, 2601 (2000)
\bibitem{Cunningham99} J. Cunningham, V. I. Talyanskii, J. M. Shilton, M. Pepper, M. Y. Simmons, and D. A. Ritchie, Phys. Rev. B 60, 4850 (1999)
\bibitem{Wei01} Y. Wei, J. Wang, H. Guo, and Ch. Roland, Phys. Rev. B 64, 115321 (2001)

\end{thebibliography}
\end{document}